\documentclass{emulateapj}
\usepackage{graphics,graphicx,rotating}
\usepackage[colorlinks,allcolors=blue]{hyperref}
\usepackage[all]{hypcap} 
\usepackage{xspace}
\usepackage{amssymb}
\usepackage{amsmath}
\usepackage{epstopdf}
\usepackage[titletoc]{appendix}
\usepackage{subfigure}
\usepackage{xcolor}
\usepackage{makecell}
\usepackage{tabularx,ragged2e}
\usepackage{natbib}
\UseRawInputEncoding
\begin{document}

\title{Ultramassive black holes in the most massive galaxies: $M_{\rm BH}-\sigma$ versus $M_{\rm BH}-R_{\rm b}$ }  

\shorttitle{Ultramassive black holes in the most massive galaxies }
\shortauthors{Dullo, Gil de Paz \& Knapen} 
\author{\color{blue}{Bililign T. Dullo,$^{1}$ Armando Gil de Paz,$^{1}$ and Johan H. Knapen$^{2,3}$ }}
\affil{\altaffilmark{1} Departamento de F\'isica de la Tierra y Astrof\'isica,
  Instituto de F\'isica de Part\'iculas y del Cosmos IPARCOS,
  Universidad Complutense de Madrid, E-28040 Madrid, Spain;
  bdullo@ucm.es}
\affil{\altaffilmark{2}Instituto de Astrof\'isica de Canarias, V\'ia
  L\'actea S/N, E-38205 La Laguna, Spain}
\affil{\altaffilmark{3}Departamento de Astrof\'isica, Universidad de
  La Laguna, E-38206 La Laguna, Spain}

\begin{abstract}

  We investigate the nature of the relations between black hole (BH)
  mass ($M_{\rm BH}$) and the central velocity dispersion ($\sigma$)
  and, for \mbox{core-S\'ersic} galaxies, the size of the depleted
  core ($R_{\rm b}$).  Our sample of 144 galaxies with dynamically
  determined $M_{\rm BH}$ encompasses 24 \mbox{core-S\'ersic}
  galaxies, thought to be products of gas-poor mergers, and reliably
  identified based on high-resolution {\it HST} imaging.  For
  \mbox{core-S\'ersic} galaxies---i.e., combining \mbox{normal-core}
  ($R_{\rm b} < 0.5 $ kpc) and \mbox{large-core} galaxies
  ($R_{\rm b} \ga 0.5$ kpc), we find that $M_{\rm BH}$ correlates
  remarkably well with $R_{\rm b}$ such that
  \mbox{$M_{\rm BH} \propto R_{\rm b}^{1.20 \pm 0.14}$} (rms scatter
  in \mbox{{log $M_{\rm BH}$} of $\Delta_{\rm rms} \sim 0.29$ dex}),
  confirming previous works on the same galaxies except three new
  ones. Separating the sample into S\'ersic, \mbox{normal-core} and
  \mbox{large-core} galaxies, we find that S\'ersic and
  \mbox{normal-core} galaxies jointly define a single log-linear
  \mbox{$M_{\rm BH}-\sigma$} relation
  $M_{\rm BH} \propto \sigma^{ 4.88 \pm 0.29}$ with
  \mbox{$\Delta_{\rm rms} \sim 0.47$ dex}, however, at the high-mass
  end \mbox{large-core} galaxies (four with measured $M_{\rm BH}$) are
  offset upward from this relation by\mbox{
    ($2.5-4) \times \sigma_{\rm s}$}, explaining the previously
  reported steepening of the \mbox{$M_{\rm BH}-\sigma$} relation for
  massive galaxies. \mbox{Large-core} spheroids have magnitudes
  $M_{V} \la -23.50$ mag, half-light radii $R_{\rm e} \ga 10$ kpc and
  are extremely massive $M_{*} \ga 10^{12}M_{\sun}$. Furthermore,
  these spheroids tend to host ultramassive BHs
  ($M_{\rm BH} \ga 10^{10}M_{\sun}$) tightly connected with their
  $R_{\rm b}$ rather than $\sigma$. The less popular
  \mbox{$M_{\rm BH}- R_{\rm b}$} relation exhibits $\sim$ 62\% less
  scatter in \mbox{log $M_{\rm BH}$} than the
  \mbox{$M_{\rm BH}- \sigma$} relations. Our findings suggest that
  \mbox{large-core} spheroids form via multiple major `dry' merger
  events involving super/ultramassive BHs, consistent with the
  flattening of the \mbox{$\sigma-L_{V}$} relation observed at
  $M_{V} \la -23.5$ mag.
\end{abstract}


\section{Introduction}

Observational works on nearby galaxies in the last 25 yrs have
revealed that black holes (BHs) in a mass range
\mbox{$M_{\rm BH} \sim 10^{6}-10^{10} M_{\sun}$} reside at the centers
of all massive elliptical galaxies and massive bulges of disk galaxies
\citep{1995ARA&A..33..581K,1998AJ....115.2285M,2005SSRv..116..523F,2011Natur.480..215M,2016Natur.532..340T,2019ApJ...887..195M}. The
BH can have a major impact on the central and large-scale properties
of its host galaxy.  Luminous ($M_{V} \la -21.50 \pm 0.75$ mag)
core-S\'ersic galaxies contain partially depleted stellar cores, a
flattening in the inner stellar light distributions, that are thought
to be generated by the scouring action of inspiraling supermassive
black hole (SMBH) binaries formed in major dry merger events. As the
inspiraling SMBH binary sinks to the center of the merger remnant, it
transfers orbital angular momentum to the surrounding stars. The
gravitation slingshot ejection of the inner stars by this decaying
SMBH binary creates the central light deficit, i.e., the flattened
core
\citep[e.g.,][]{1980Natur.287..307B,1991Natur.354..212E,2001ApJ...563...34M,
  2006ApJ...648..976M,2012ApJ...744...74G,2013ApJ...773..100K,2015ApJ...810...49V,2018ApJ...864..113R,2020arXiv201104663N}.
The light profiles of core-S\'ersic galaxies, which break from a steep
outer S\'ersic profile to a flattened core, are well described using
the core-S\'ersic model \citep{2003AJ....125.2951G}. This model
enables the core sizes of core-S\'ersic galaxies to be measured by its
break radius $R_{\rm b}$
\citep[e.g.,][]{2003AJ....125.2951G,2006ApJS..164..334F,
  2012ApJ...755..163D,2014MNRAS.444.2700D,2013ApJ...768...36D,2019ApJ...886...80D,
  2013AJ....146..160R}. Recently, \citet{2019ApJ...886...80D} revealed
two types of core-S\'ersic galaxies: ``normal-core'' (i.e.,
$R_{\rm b} < $ 0.5 kpc) and ``large-core'' (i.e., $R_{\rm b} >$ 0.5
kpc) galaxies.

The SMBH masses have been found to correlate with several host galaxy
properties including central stellar velocity dispersion ($\sigma$,
\citealt{2000ApJ...539L...9F,2000ApJ...539L..13G,2011MNRAS.412.2211G,2013ApJ...764..184M}),
luminosity ($L_{\rm sph}$, e.g.,
\citealt{1995ARA&A..33..581K,2002MNRAS.331..795M,2013ApJ...764..151G})
and mass ($M_{\rm sph}$, e.g.,
\citealt{1998AJ....115.2285M,2003ApJ...589L..21M,2004ApJ...604L..89H,2019ApJ...876..155S}). Later
studies found that a tight relation exists between SMBH masses and the
sizes of the depleted cores for core-S\'ersic galaxies ($R_{\rm b}$,
e.g.,
\citealt{2007ApJ...662..808L,2013AJ....146..160R,2014MNRAS.444.2700D,
  2016Natur.532..340T,2019ApJ...886...80D}).
\citet{2020ApJ...898...83D} discover correlations between SMBH mass
and the host galaxy total UV-[3.6] color. In this new (SMBH
mass)-color diagram, early- and late-type galaxies define a red and
blue sequence, respectively.  These local SMBH scaling relations have
generated great interest as they are believed to indicate a strong
evolutionary coupling between the growth of a SMBH and the buildup of
its host galaxy (see recent reviews by
\citealt{2013ARA&A..51..511K,2016ASSL..418..263G}) governed by perhaps
active galactic nucleus (AGN) feedback
\citep[e.g.,][]{1998A&A...331L...1S, 1999MNRAS.308L..39F,
  2005Natur.435..629S,2005Natur.433..604D,2006MNRAS.365...11C,
  2006ApJS..163....1H} or hierarchical merging processes
\citep[e.g.,][]{2007ApJ...671.1098P,2011ApJ...734...92J}. The
expectation is that the slopes, strength, level of scatter and
substructures of such relations can yield clues to the underlying
mechanism that establishes supposed coupling between SMBHs and host
galaxies.

It seems reasonable that the central BH is more intimately related to
the size of the depleted core it created than the stellar velocity
dispersion.  However, the $M_{\rm BH}-\sigma$ relation is regarded as
the most fundamental of all the SMBH scaling relations due to its
tightness and small scatter \citep[e.g.,][$\sim0.3$ dex in the
\mbox{log $M_{\rm BH}$}]{2002ApJ...574..740T}, which are claimed to be
comparable to those of the scaling relations involving $M_{\rm BH}$
and two spheroid parameters (e.g., $\sigma, L_{\rm sph}$ and
half-light radii $R_{\rm e}$), defining a ``BH fundamental plane''
\citep[e.g.,][]{2012MNRAS.419.2497B,2016ApJ...818...47S,2016MNRAS.460.3119S,2016ApJ...831..134V,2018MNRAS.473.5237K,2019MNRAS.485.1278S,2019MNRAS.490..600D}.
Therefore, the relation is commonly used for estimating the SMBH
masses of nearby galaxies when direct SMBH mass measurements are not
available. This is particularly true for the brightest and most
massive galaxies with very faint central surface brightness (e.g.,
\citealt{2019ApJ...886...80D}) that render the modeling stellar or
ionized gas kinematics for the dynamical SMBH mass measurements
challenging. In contrast, the $M_{\rm BH}-R_{\rm b}$ relation has been
largely overlooked, despite displaying a similar level of scatter as
the $M_{\rm BH}-\sigma$ relation
\citep[e.g.,][]{2013AJ....146..160R,2014MNRAS.444.2700D,
  2016Natur.532..340T,2019ApJ...886...80D}. This is mainly because the
offset nature of the most massive galaxies (at the high-mass end) from
the mean $M_{\rm BH}-\sigma$ relation is still not well established
and also the $M_{\rm BH}-R_{\rm b}$ relation can only be applied to
massive core-S\'ersic spheroids believed to be end-products of
gas-poor major mergers
\cite[e.g.,][]{1997AJ....114.1771F,2009ApJS..181..486H}. At low and
intermediate luminosities ($M_{V} \ga -21.0$ mag), gas-rich processes
are commonly thought to produce S\'ersic spheroids with no depleted
cores and with $\sigma \la 180$ km s$^{-1}$, S\'ersic index $n < 3$,
and $M_{\rm BH} \la 10^{8} M_{\sun}$
\citep[e.g.,][]{2006ApJS..164..334F,2009ApJS..181..135H,2012ApJ...755..163D,2013ApJ...764..151G,2016MNRAS.462.3800D,2019ApJ...871....9D,2020ApJ...898...83D}.

The majority of dynamical BH masses measured to date are SMBHs
residing in massive (but not the most mass) galaxies. However, recent
studies obtained ultramassive black holes (\mbox{UMBHs,
  $M_{\rm BH} \ga 10^{10} M_{\sun}$}) in the most massive galaxies,
which offset upward and toward large $M_{\rm BH}$ in the
$M_{\rm BH}-\sigma$ diagrams
\citep[e.g.,][]{2011Natur.480..215M,2013ApJ...764..184M,2016Natur.532..340T,2018MNRAS.474.1342M,2019ApJ...886...80D,2019ApJ...887...10S}. This
is consistent with \citet{2007ApJ...662..808L} who using a galaxy
sample with $M_{\rm BH} \la 3 \times 10^{9} M_{\sun}$ speculated that
SMBH masses for the brightest cluster galaxies (BCGs) would be
overmassive relative to the expectation from the high mass end of the
\mbox{$M_{\rm BH}-\sigma$} relation, but they would be in better
agreement with those from \mbox{$M_{\rm BH}-L$} relation (see also
\citealt{2013ApJ...768...29V,2019ApJ...886...80D}). The
$M_{\rm BH}-\sigma$ relation predicts SMBH masses for the most massive
galaxies (i.e., \mbox{$\sigma \sim 300 -390$ km s$^{-1}$},
\citealt{2007ApJ...662..808L,2007AJ....133.1741B}) of
$M_{\rm BH} \la 5 \times 10^{9} M_{\sun}$, and these values
underestimate the actual black hole mass by up to a factor of 40
\citep{2019ApJ...886...80D}, whereas predicted $M_{\rm BH}$ from the
$M_{\rm BH}-L$ relation can exceed $M_{\rm BH} \sim 10^{10} M_{\sun}$
(\citealt{2007ApJ...662..808L,2019ApJ...886...80D}). Indeed,
\citet{2015Natur.518..512W} found a quasar at a redshift of z $\sim$
6.3 powered by an UMBH with
\mbox{$M_{\rm BH} \sim 1.2\times 10^{10} M_{\sun}$}, and such BHs have
been identified at the centers of extremely massive
($M_{*} \ga 10^{12}M_{\sun}$) present-day galaxies
\citep[e.g.,][]{2011Natur.480..215M,2016Natur.532..340T,2019ApJ...887..195M}.

Separating core-S\'ersic galaxies into normal-core and large-core
galaxies, our findings \citep{2019ApJ...886...80D} suggested that the
offset evident at the high-mass end of the $M_{\rm BH}-\sigma$
relation is due to large-core spheroids with
\mbox{$M_{V} \la -23.50 \pm 0.10$ mag}, $M_{*} \ga 10^{12}M_{\sun}$
and $R_{\rm e} \ga 10$ kpc. It worth noting such spheroids are not
simple high-mass extensions of the relatively less massive normal-core
spheroids ($ M_{*} \sim 8\times 10^{10} - 10^{12}M_{\sun}$).  A key
observation in \citet{2019ApJ...886...80D} was that a single
$M_{\rm BH}-R_{\rm b}$ relation holds across the full mass range of
core-S\'ersic spheroids confirming \citet{2007ApJ...662..808L,2013AJ....146..160R,2014MNRAS.444.2700D}, but this conclusion was based on a small
sample of 11 core-S\'ersic galaxies with direct SMBH mass
measurements; 3/11 galaxies are large-core galaxies. Establishing this
tight $M_{\rm BH}-R_{\rm b}$ relation, as well as the offset at the
high-mass end of the $M_{\rm BH}-\sigma$ relation being due to
large-core galaxies, requires a reasonably large sample of galaxies
with dynamical SMBH mass measurements.

In this work, we build upon \citet{2019ApJ...886...80D} and
investigate the nature of the $M_{\rm BH}-R_{\rm b}$ and
$M_{\rm BH}-\sigma$ relations using 24 core-S\'ersic galaxies with published 
direct SMBH mass determinations. We additionally include 27
core-S\'ersic with predicted SMBH masses and robust break radii to
further investigate the the BH scaling relations. Our full sample of
144 galaxies with dynamically measured SMBH masses, together with the
27 core-S\'ersic galaxies with predicted $M_{\rm BH}$, used to
directly explore substructures in the $M_{\rm BH}-\sigma$ diagram is
described in \mbox{Section~\ref{Sec_2_data}}. We then discuss the
linear regression methods employed in \mbox{Section~\ref{Sec3.1}}.  In
\mbox{Section~\ref{Sec3.2}}, we show that the $M_{\rm BH}-R_{\rm b}$
relation for core-S\'ersic galaxies is stronger and has less scatter
than the core-S\'ersic $M_{\rm BH}-\sigma$ relation. We reveal the
steepening at the high-mass end of the $M_{\rm BH}-\sigma$ relation in
Section~\ref{Sec3.3} and go on to discuss how underestimated BH masses
of large-core galaxies give rise to offsets in the
$R_{\rm b}-M_{\rm BH}$ diagrams in \mbox{Section~\ref{Sec3.4}}. In
\mbox{Section~\ref{Bhgrowth}}, we discuss pathways for the growth of
black holes in large-core, normal-core and S\'ersic
galaxies. \mbox{Section~\ref{Conc}} summarizes our main conclusions.


\section{SAMPLE AND DATA}\label{Sec_2_data}
  
We use the sample of 41 core-S\'ersic galaxies and their robust break
radii from \citet[][see also
\citealt{2014MNRAS.444.2700D,2017MNRAS.471.2321D}]{2019ApJ...886...80D}.
Of these 41 galaxies, 14 have dynamically determined (henceforth
``direct'') SMBH masses. We exclude the measured black hole masses for
two \citet{2014MNRAS.444.2700D} core-S\'ersic galaxies
(\mbox{NGC~3706}, \citealt{2014ApJ...781..112G} and \mbox{NGC~5419},
\citealt{2016MNRAS.462.2847M}) with an inner stellar ring and two
compact nuclear point sources, respectively
\citep{2012ApJ...755..163D}.  These two black hole masses are
potentially less secure due to the galaxies' complex nuclear
structures. We add here 10 other core-S\'ersic galaxies with direct
SMBH masses and robust core-S\'ersic break radii\footnote{We convert
  the geometric-mean break radii from \cite{2013AJ....146..160R} into
  semi-major axis radii using the galaxies' ellipticity values at the
  break radii.}  from \cite{2013AJ....146..160R}. This results in a
final compilation sample of 51 (13 large-core and 38 normal-core)
core-S\'ersic galaxies having measured break radii.

\begin{figure}
\hspace*{1.0919cm}   
\vspace*{0.109142562932599cm}   
\includegraphics[angle=270,scale=0.5853]{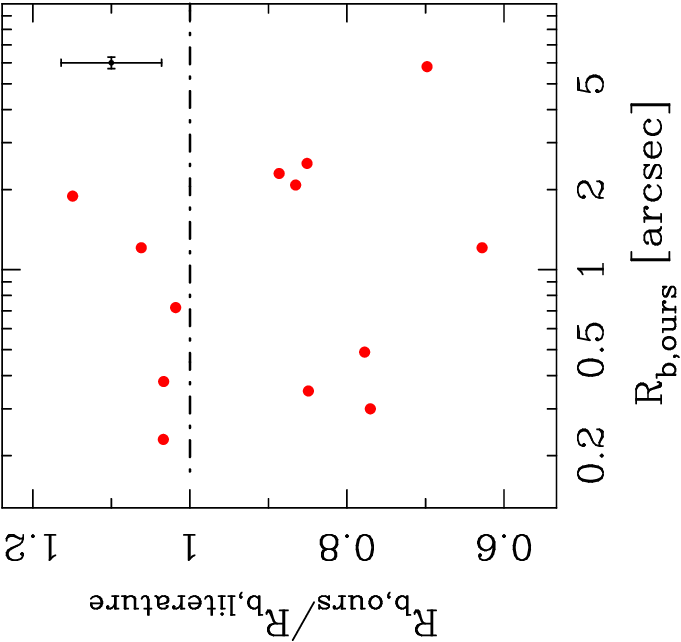}
\caption{Comparison of our core-S\'ersic break radii \citep{2014MNRAS.444.2700D,2019ApJ...886...80D}  with previous core-S\'ersic break radii values from \citet{2013AJ....146..160R,2016Natur.532..340T} after converting  their geometric-mean break radii  into semi-major axis radii using the galaxies' ellipticity values at the break radii. The typical uncertainty on $R_{\rm b}$ is \mbox{$\sim5$ \%} \cite[e.g.,][]{2013AJ....146..160R,2019ApJ...886...80D}.  A representative error bar  is shown.}
\label{Fig0}
 \end{figure}

Of the 24 core-S\'ersic galaxies with direct SMBH mass measurements in
our final sample (Table~\ref{Table0}), 21 are in common with
\citet{2016Natur.532..340T}, see also
\citet{2013AJ....146..160R}. \citet{2016Natur.532..340T} used their
break radius for NGC~1600 and the break radii from
\cite{2013AJ....146..160R} for the remaining 20 core-S\'ersic galaxies
in their sample.  The break radii in this paper, in contrast, are from
our own analyses
\citep{2014MNRAS.444.2700D,2017MNRAS.471.2321D,2019ApJ...886...80D}
except for the 10 galaxies taken from \cite{2013AJ....146..160R}, see
Table~\ref{Table0}. Fig.~\ref{Fig0} shows a comparison of our
core-S\'ersic break radii with the core-S\'ersic break radii values
from \cite{2013AJ....146..160R,2016Natur.532..340T} for 13 overlapping
galaxies. For four galaxies, our break radii agree with theirs within
\mbox{10 \%}, whereas for the remaining 9 galaxies there is a
discrepancy of \mbox{$\sim13-59$ \%}.  The typical uncertainty
associated with $R_{\rm b}$ is \mbox{$\sim5$ \%}
\cite[e.g.,][]{2013AJ....146..160R,2019ApJ...886...80D}. The two
galaxies where our break radii disagree most ($\sim$ \mbox{43-59 \%})
with those of \cite{2013AJ....146..160R} are the two brightest
galaxies in the Virgo cluster \mbox{NGC 4472} and \mbox{NGC
  4486}.

The identification of the partially depleted cores for the large-core
and normal-core galaxies is based on detailed structural decomposition
of the galaxies' {\it HST} surface brightness profiles, fitting the
core-S\'ersic model to the spheroidal components
(\citealt{2019ApJ...886...80D,2013AJ....146..160R}).

In order to construct the $M-\sigma$ relations shown in this paper, we
used \citet[see their Table~2]{2016ApJ...831..134V} as main
reference. He tabulated a large compilation of 245 measured SMBH
masses.  Excluding his galaxies in common with
\citet{2013AJ....146..160R,2019ApJ...886...80D}, we consider secure
SMBH masses measured based on stellar and gas dynamics
\citep{2016ApJ...831..134V}.  To avoid issues of potential
inconsistency arising from using upper limits and BH masses obtained
via different methods, we excluded SMBH mass upper limits, and SMBH
masses from megamasers and reverberation mapping measurements. This
leaves us with a sample of 121 dynamically determined SMBH masses and
the majority of these galaxies are S\'ersic galaxies, while a small
fraction of them with $M_{\rm BH }\ga 2 \times 10^{9} M_{\sun}$ and
\mbox{$\sigma \ga 280 $ km s$^{-1}$} are likely core-S\'ersic
galaxies. Homogenized mean central velocity dispersions ($\sigma$) for
the sample galaxies were obtained from
HyperLeda\footnote{http://leda.univ-lyon1.fr}
\citep{2014A&A...570A..13M}. We adopt a conservative upper limit
uncertainty of 10\% on $\sigma$ after comparing the HyperLeda
individual velocity dispersion measurements and mean homogenized
values for about 100 sample galaxies.

\begin{table}
\caption{Core-S\'ersic galaxies with measured black hole masses.}
\label{Table0}
\begin {minipage}{220mm}
\setlength{\tabcolsep}{0.3863in}   
\begin{tabularx}{0.39\textwidth}{@{}lllcccccccccccccccc@{}}\hline
\hline
Galaxy&$R_{\rm b}$ (kpc)&log~($M_{\rm BH}/M_{\sun}$)\\
(1)&(2)&(3)&\\
\multicolumn{1}{c}{} \\     
\toprule
IC~1459 &0.108 [1]	&	9.45$^{+0.14}_{-0.24}$  [1]  \\
NGC~0584  &0.021 [2]& 8.15$^{+0.13}_{-0.19}$  [3]   \\ 
NGC~1399   &0.202 [2] & 9.07$^{+0.7}_{-0.46}$  [2]  \\ 
NGC~1407 &0.276 [1]&	9.65$^{+0.08}_{-0.04}$ [1] 	\\
NGC~1550 &0.300 [1] & 9.57$^{+0.05}_{-0.05}$ [1] \\	
NGC~1600&0.650  [4] &10.23$^{+0.04}_{-0.04}$ [4] \\
NGC~3091&0.169 [1]&	9.56$^{+0.01}_{-0.03}$ [1]\\
NGC~3379  &0.102 [2]& 8.60$^{+0.10}_{-0.13}$ [2]  \\ 
NGC~3608  &0.024 [2]&8.30$^{+0.19}_{-0.16}$  [2] \\ 
NGC~3842  &0.315 [2]& 9.98$^{+0.12}_{-0.14}$ [2]\\ 
NGC~4261 &0.198 [1]& 	8.72$^{+0.08}_{-0.10}$ [1]	\\
NGC~4291&0.036  [2]	 & 8.52$^{+0.11}_{-0.62}$  [2]  \\ 
NGC~4374  &0.139 [1] &8.96$^{+0.05}_{-0.04}$ [1]\\
NGC~4472  &0.108 [2] &9.36$^{+0.04}_{-0.02}$ [1] \\
NGC~4486 & 0.640 [4] &$9.76^{+0.03}_{-0.03}$ [4] \\
NGC~4552 	 &0.017 [2]&8.67$^{+0.04}_{-0.05}$ [2]  \\ 
NGC~4649	 &0.241 [2]&9.67$^{+0.08}_{-0.10}$ [2]  \\ 
NGC~4889&0.860 [4] &10.30$^{+0.25}_{-0.62}$ [4] \\
NGC~5328&0.271 [1]&	9.67$^{+0.08}_{-0.23}$ [1]\\
NGC~5516&0.178 [1]&	9.52$^{+0.03}_{-0.04}$ [1]	\\
NGC~6086	&0.357 [1]	 &	9.56$^{+0.17}_{-0.16}$ [1]\\	
NGC~7768&0.164	 [1]&		9.11$^{+0.14}_{-0.16}$  [1]	\\
NGC~5813	 & 0.051 [2]& 8.83$^{+0.04}_{-0.05}$ [2]   \\ 
NGC~7619	 & 0.109  [2]& 9.36$^{+0.06}_{-0.12}$ [5]  \\  
\mbox{Holm 15A}? &  2.800 [6] &10.60$^{+0.08}_{-0.10}$ [6]\\
\hline
\end{tabularx}
\end {minipage}
Note. Col.\ (1) galaxy name. Col.\ (2) break radius. Col.\ (3) SMBH mass.
Sources.  [1]=\citet[][and references therein]{2013AJ....146..160R}; [2]=\citet[][and references therein]{2014MNRAS.444.2700D}; [3]=\citet{2019AA...625A..62T},
[4]=\citet[][and references therein]{2019ApJ...886...80D}; [5]=\citet{2013AJ....146...45R}; [6]=\citet{2019ApJ...887..195M}. A `?' is used to indicate that the identification of a  depleted core in \mbox{Holm 15A} is uncertain.
\end{table}

\begin{table}
\caption{Large-core galaxy data}
\label{Table2}
\begin {minipage}{220mm}
\setlength{\tabcolsep}{0.03818in}   
\begin{tabularx}{0.39\textwidth}{@{}lllcccccccccccccccc@{}}\hline
\hline
Galaxy&&log~($M_{\rm BH}/M_{\sun}$)&&$M_{\rm def}/M_{\rm BH}$\\
&($\sigma$-based)&(L-based)&($R_{\rm b}$-based)&($\sigma$/$L_{\rm sph}$/$R_{\rm b}$)\\
(1)&(2)&(3)&(4)&(5)&\\
\multicolumn{1}{c}{} \\           
\toprule
NGC~4874      & 8.91$^{+0.42}_{-0.42}$&8.98$^{+0.34}_{-0.34}$& 10.54$^{+0.45}_{-0.45}$&79.8/67.6/1.9\\ 
NGC~6166      &9.14$^{+0.43}_{-0.43}$&10.47$^{+0.47}_{-0.47}$& 10.68$^{+0.46}_{-0.46}$&117.3/5.5/3.4 \\
4C+74.13  &8.61$^{+0.41}_{-0.41}$&10.20$^{+0.43}_{-0.43}$&  10.71$^{+0.46}_{-0.46}$&157.0/4.1/1.3\\
A0119    &9.04$^{+0.43}_{-0.43}$&10.41$^{+0.46}_{-0.46}$&10.08$^{+0.42}_{-0.42}$ &13.5/0.6/1.2\\  
A2029    &9.68$^{+0.47}_{-0.47}$&  9.99$^{+0.40}_{-0.40}$  &10.99$^{+0.44}_{-0.44}$&77.6/50.4/4.8\\
A2147   &8.95$^{+0.42}_{-0.42}$& 9.93$^{+0.40}_{-0.40}$ &10.42$^{+0.44}_{-0.44}$&61.2/6.4/2.1\\
A2261  &9.73$^{+0.48}_{-0.48}$&10.43$^{+0.48}_{-0.48}$&10.81$^{+0.45}_{-0.45}$&  8.2/1.7/0.7\\ 
A3558    &8.70$^{+0.41}_{-0.41}$&10.89$^{+0.54}_{-0.54}$&10.43$^{+0.44}_{-0.44}$&85.7/0.6/1.7\\
A3562   &8.59$^{+0.41}_{-0.41}$&9.83$^{+0.39}_{-0.39}$&10.05$^{+0.42}_{-0.42}$&34.6/2.0/1.2\\ 
A3571   &9.31$^{+0.44}_{-0.44}$&9.79$^{+0.39}_{-0.39}$&10.44$^{+0.43}_{-0.43}$&28.0/9.3/2.1\\ 
\hline
Galaxy&&log~($M_{\rm BH}/M_{\sun}$)&&$M_{\rm def}/M_{\rm BH}$\\
&\multicolumn{8}{c}{(dynamically determined SMBH masses)}&&&\\
\hline
NGC 1600&&10.23$^{+0.04}_{-0.04}$&&2.8/2.8/2.8\\
NGC  4486&& $9.76^{+0.03}_{-0.03}$&&5.1/5.1/5.1\\
NGC  4889&& 10.3$^{+0.25}_{-0.62}$&&4.7/4.7/4.7\\
\hline
\hline
\end{tabularx}
\end {minipage}
Note. Col.\ (1) galaxy name. Cols.\ (2-4) SMBH masses 
either dynamically determined or  predicted
using the velocity dispersion $\sigma$, spheroid luminosity $L_{\rm sph}$ and
break radius $R_{\rm b}$ for  13 large-core ellipticals from \citet[][and references therein]{2019ApJ...886...80D}.  Col.\ (5) (central stellar mass deficit)-to-(SMBH mass) ratio
$M_{\rm def}/M_{\rm BH}$ \citep{2019ApJ...886...80D}. 
\end{table}

\setlength{\tabcolsep}{0.0078048090in}
\begin{table*}
\begin {minipage}{180mm}
\caption{Scaling relations } 
\label{Table1}
\begin{tabular}{@{}lllccccccccccccccccccccccccccccccc@{}}
\hline
\hline
Relation &BCES bisector fit& $\Delta_{\rm rms}$ [dex] &$\epsilon$ [dex] &$r_{\rm
                                              s}/P$-${\rm value}$&$r_{\rm
                                                            p}/P$-${\rm value}$&Sample&\\
\hline
\Xhline{2\arrayrulewidth}
 \multicolumn{8}{c}{\bf Core-S\'ersic galaxies }\\
\Xhline{2\arrayrulewidth}
$M_{\rm BH}-R_{\rm b} $
        &$\mbox{log}\left(\frac{M_{\rm BH}}{M_{\sun}}\right)= (1.20\pm
           0.14) \mbox{log}\left(\frac{R_{\rm b}}{\mbox{250 pc}}\right)$ +~($9.52 ~ \pm  0.06$)&0.29&0.33$\pm$  0.07
 &  0.90/$5.6 \times 10^{-9}$&0.88/$ 2.5\times10^{-8}$&24 [a]\\
$M_{\rm BH}-\sigma $
        &$\mbox{log}\left(\frac{M_{\rm BH}}{M_{\sun}}\right)= (10.67\pm
            4.90) \mbox{log}\left(\frac{\sigma}{\mbox{300 {\rm km s$^{-1}$}}}\right)$ +~($9.48 ~ \pm  0.11$)& 0.47&0.30$\pm$ 0.13
 
 &0.78/$2.2\times10^{-6}$&0.76/$1.3\times10^{-5}$&24 [a]\\
$M_{\rm BH}-\sigma $
        &$\mbox{log}\left(\frac{M_{\rm BH}}{M_{\sun}}\right)= (11.87\pm
             4.79) \mbox{log}\left(\frac{\sigma}{\mbox{300 {\rm km s$^{-1}$}}}\right)$ +~($9.87~ \pm  0.15$)& 0.81&0.52$\pm$ 0.14
 &0.36/$4.6 \times10^{-2}$&0.46/$10^{-2}$&34 [b]\\
\Xhline{2\arrayrulewidth}
 \multicolumn{8}{c}{\bf S\'ersic plus Core-S\'ersic galaxies }\\
\Xhline{2\arrayrulewidth}
$M_{\rm BH}-\sigma $
        &$\mbox{log}\left(\frac{M_{\rm BH}}{M_{\sun}}\right)= (4.88\pm
             0.29) \mbox{log}\left(\frac{\sigma}{\mbox{200  {\rm km s$^{-1}$}}}\right)$    +~($8.35 ~\pm  0.04$)& 0.47&0.39$\pm$0.04
 &0.85/$ 2.1\times10^{-39}$&0.85/$  1.2\times10^{-39}$&141 [c] \\
$M_{\rm BH}-\sigma $
        &$\mbox{log}\left(\frac{M_{\rm BH}}{M_{\sun}}\right)= (5.03\pm
             0.28) \mbox{log}\left(\frac{\sigma}{\mbox{200  {\rm km s$^{-1}$}}}\right)$+~($8.37 ~\pm  0.04$)& 0.48 &0.40$\pm$0.04
 &0.86/$1.1\times10^{-41}$&0.85/$4.0\times10^{-41}$&145  [d]\\
$M_{\rm BH}-\sigma $
        &$\mbox{log}\left(\frac{M_{\rm BH}}{M_{\sun}}\right)= (5.76\pm
             0.37) \mbox{log}\left(\frac{\sigma}{\mbox{200  {\rm km s$^{-1}$}}}\right)$ +~($8.45 ~\pm  0.05$)& 0.60&0.45$\pm$0.04
 &0.85/$ 6.1\times10^{-44}$&0.82/$ 7.1\times10^{-39}$&155 
                                                       [e]\\

\hline  
\end{tabular} 
Note. The different columns represent: the BH scaling relations, rms
scatter in the vertical log $M_{\rm BH}$ direction
($\Delta_{\rm rms}$), intrinsic scatter from the Bayesian {\sc
  linmixerr} fits ($\epsilon$), and the Spearman and Pearson
correlation coefficients ($r_{\rm s}$ and $r_{\rm p}$, respectively)
and the associated probabilities. Sample: [a] 24 core-S\'ersic
galaxies with dynamically determined $M_{\rm BH}$ (i.e., 11
normal-core galaxies from \citealt{2014MNRAS.444.2700D}, 10
normal-core elliptical galaxies from \citealt{2013AJ....146..160R}
plus 3 large-core elliptical galaxies from
\citet{2019ApJ...886...80D}; [b] 34 core-S\'ersic galaxies (i.e., 24
galaxies [a] plus 10 large-core galaxies with SMBH masses predicted
using the $M_{\rm BH}-R_{\rm b}$ relation, Table~\ref{Table2}; [c] 142
non-(large-core) galaxies (i.e., 21 normal-core galaxies from [a] plus
121 galaxies with dynamically determined SMBH mass measurements from
\citet{2016ApJ...831..134V} that are not in common with
\citet{2013AJ....146..160R,2014MNRAS.444.2700D,2019ApJ...886...80D};
[d] 145 galaxies (i.e., 142 galaxies [c] plus 3 large-core ellipticals
from \citet{2019ApJ...886...80D}; [e] 155 galaxies (i.e., 145 galaxies
[d] plus 10 large-core galaxies with $R_{\rm b}$-based, predicted SMBH
masses, Table~\ref{Table2}).
\end{minipage}
\end{table*}

\subsection{Predicted black hole masses for core-S\'ersic galaxies}

\citet[][and references therein]{2019ApJ...886...80D} presented
predicted SMBH masses for the sample of 27 core-S\'ersic galaxies with
no direct SMBH masses. We made use of the
\citet[their Table 3]{2013ApJ...764..151G} non-barred
\mbox{$M_{\rm BH}-\sigma$} relation to predict the
\mbox{$\sigma$--based} SMBH masses. The predicted
\mbox{$L_{\rm sph}$--based} SMBH masses were based on the
near-linear \citet[their Table 3]{2013ApJ...764..151G} $B$-band
core-S\'ersic $M_{\rm BH}-L$ relation transformed here into the
$V$-band using $B-V = 1.0$ (\citealt{1995PASP..107..945F}).

In an effort to better investigate the galaxy-black hole co-evolution
in the most massive galaxies, we present SMBH masses predicted using
the velocity dispersion $\sigma$, spheroid luminosity $L_{\rm sph}$
and break radius $R_{\rm b}$ for the 10 large-core galaxies in our
sample with no direct SMBH masses (Table~\ref{Table2}). All the
large-core galaxies in our sample (Table~\ref{Table2}) are classified
as BCGs except for three galaxies (NGC 1600, NGC 4486 and NGC
4874). The elliptical galaxy \mbox{NGC 1600} is the brightest member
of the poor \mbox{NGC 1600} group. The giant ellipticals \mbox{NGC
  4486} and \mbox{NGC 4874} are second brightest galaxies residing at
the heart of the Virgo cluster and Coma cluster, respectively.

\section{Results and discussion}

\begin{figure*}
\hspace*{1.90919cm}   
\vspace*{0.109142562932599cm}   
\includegraphics[angle=270,scale=0.739]{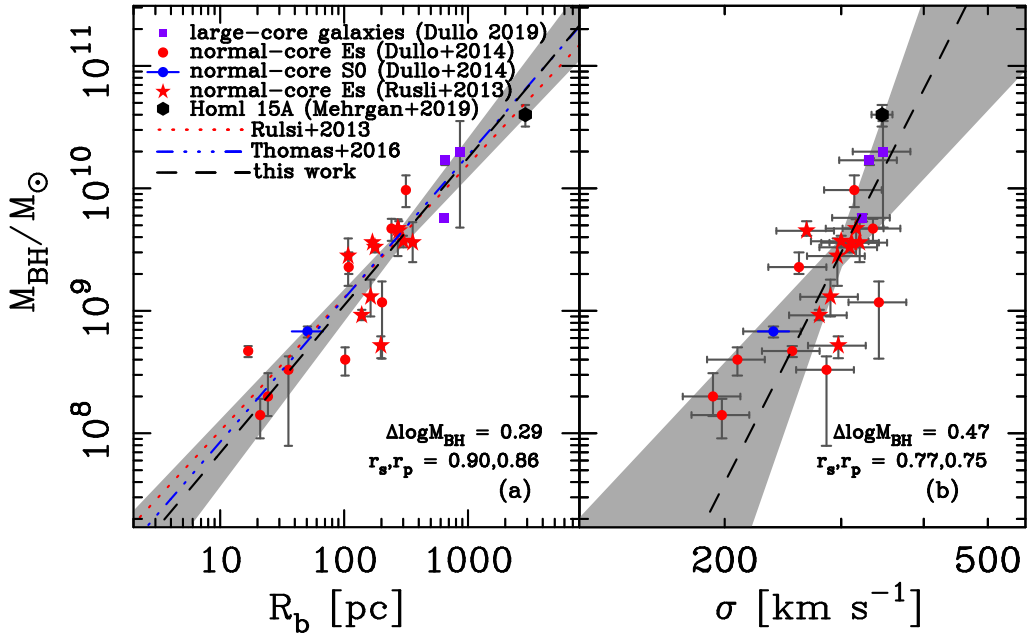}
\caption{SMBH mass ($M_{\rm BH}$) plotted as a function of (a) the
  core-S\'ersic break radius ($R_{\rm b}$) and (b) central velocity
  dispersion ($\sigma$) for a sample of 24 core-S\'ersic galaxies with
  dynamically determined values of $M_{\rm BH}$. The dashed lines are
  our symmetric {\sc bces} bisector regression fits
  (Table~\ref{Table1}). We did not include Holm 15A (filled hexagon,
  \citealt{2019ApJ...887..195M}) in the regression analyses (see the
  text for details). The shaded regions show the 1$\sigma$ uncertainty
  for the regression fits.  Filled red circles and blue disk symbol
  correspond to the 10 normal-core (i.e., $R_{\rm b} < 0.5$ kpc)
  elliptical galaxies and the 1 normal-core S0 galaxy from
  \citet[their Table~2]{2014MNRAS.444.2700D}, while filled purple
  boxes indicate the 3 large-core (i.e., $R_{\rm b} > 0.5$ kpc)
  elliptical galaxies from \citet{2019ApJ...886...80D}. The 10
  normal-core elliptical galaxies from \citet{2013AJ....146..160R} are
  denoted by filled red stars. We show the uncertainties on
  $R_{\rm b}$, but they are smaller than the symbol sizes. The rms
  scatter in the vertical log $M_{\rm BH}$ direction
  ($\Delta_{\rm rms}$) and the Spearman and Pearson correlation
  coefficients ($r_{\rm s}$ and $r_{\rm p}$, respectively) are shown
  at the bottom of the panels. The dotted and dashed-dotted lines
  represent the $M_{\rm BH}-R_{\rm b}$ relations from
  \citet{2013AJ....146..160R} and \citet{2016Natur.532..340T}.}
\label{Fig1}
 \end{figure*}

\begin{figure}
\hspace*{-.205250807091cm}   
\vspace*{.109142562932599cm}   
\includegraphics[angle=270,scale=0.83729]{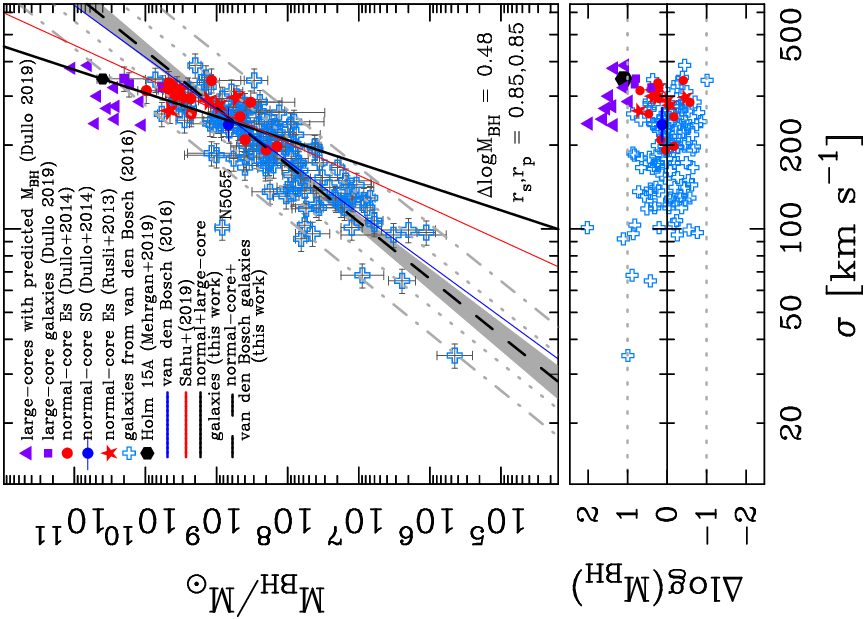}
\caption{Correlation between SMBH mass ($M_{\rm BH}$) and stellar
  velocity dispersion ($\sigma$). Similar to Fig.~\ref{Fig1}(b) but
  here we also show the 121 galaxies with dynamically determined SMBH
  masses from \citet{2016ApJ...831..134V} that are not in common with
  \citet{2013AJ....146..160R,2014MNRAS.444.2700D,2019ApJ...886...80D}, open
  crosses and 10 large-core galaxies with SMBH masses predicted using the
  \mbox{$M_{\rm BH}-R_{\rm b}$} relation (filled triangles,
  Table~\ref{Table1} and \citet[][his Table~5]{2019ApJ...886...80D}. The dashed line
  represents our symmetric {\sc bces} bisector fit to the
  ($M_{\rm BH}, \sigma$) data for the composite sample of 141
  galaxies---121 galaxies from \citet{2016ApJ...831..134V}, 10
  normal-core galaxies \citep{2014MNRAS.444.2700D} and 10 normal-core
  ellipticals \citep{2013AJ....146..160R}. The shaded region shows the
  1$\sigma$ uncertainty for the fit. The dotted and dashed-dotted
  lines delineate one and two times the measured vertical rms scatter
  in the \mbox{log $M_{\rm BH}$} direction ($\Delta_{\rm rms} = 0.48$
  dex); the residual profile about the fit is given in the lower
  panel. The solid black line is the {\sc bces} bisector fit to the
  31 (normal- and large-core) core-S\'ersic  galaxies (see Table~1). The solid blue line is the
  $M_{\rm BH}-\sigma$ relation found by \citet{2016ApJ...831..134V}
  for their full sample of 230 galaxies, while the solid red line is
  the core-S\'ersic $M_{\rm BH}-\sigma$ relation reported by
  \citet{2019ApJ...887...10S}. }
\label{Fig2}
 \end{figure}

 \begin{figure}
\hspace*{-.14cm}   
\includegraphics[angle=270,scale=0.58]{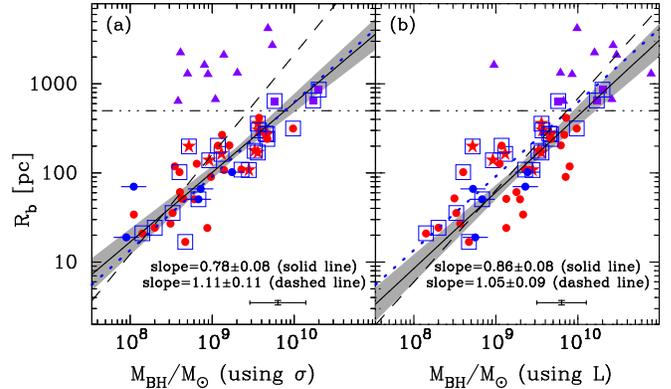}
\caption{$R_{\rm b}-M_{\rm BH}$ diagrams for 51 core-S\'ersic
  galaxies,  symbolic representations are as in
  Fig.~\ref{Fig2}. Similar to Fig.~\ref{Fig1}(a), but here we include
  27 (17
  normal-core plus 10 large-core) core-S\'ersic galaxies with SMBH
  masses that are predicted using the \citet{2013ApJ...764..151G}
  non-barred $M_{\rm BH}-\sigma$ relation (a) and their $B$-band
  core-S\'ersic $M_{\rm BH}-L$ relation (b).  The dotted line
  represents the $R_{\rm b}-M_{\rm BH}$ relation for our full sample
  of 24 core-S\'ersic galaxies (enclosed in boxes) with measured SMBH
  masses (see Fig.~\ref{Fig1}a and Table~\ref{Table1}). The solid
  lines are the symmetric {\sc ols} bisector regressions for 41
  core-S\'ersic galaxies (excluding the 10 large-core galaxies with
  predicted SMBH masses) and the shaded regions show the associated
  1$\sigma$ uncertainties. The dashed lines are the symmetric {\sc
    ols} bisector fits to the full sample of 51 core-S\'ersic
  galaxies. Large-core spheroids at the high-mass end lie above the best-fitting
  $R_{\rm b}-M_{\rm BH}$ relation because the $M_{\rm BH}-\sigma$ and
  core-S\'ersic $M_{\rm BH}-L$ relations underestimate SMBH masses in
  extremely massive galaxies. }
\label{Fig3} 
 \end{figure}

\subsection{Linear regressions}\label{Sec3.1}

We performed linear regression fits to the
(\mbox{$M_{\rm BH}, R_{\rm b}$}) and ($M_{\rm BH},\sigma$) data sets
(Figs.~\ref{Fig1} and \ref{Fig2}) using two regression techniques: the
Bivariate Correlated Errors and intrinsic Scatter ({\sc bces}) code
(\citealt{1996ApJ...470..706A}) and the Bayesian linear regression
routine ({\sc linmix\_err}, \citealt{2007ApJ...665.1489K}). The {\sc
  bces} routine \citep{1996ApJ...470..706A} was implemented in our
work using the python module by \citet{2012Sci...338.1445N}. Both the
{\sc bces} and \mbox{\sc linmix\_err} methods take into account the
intrinsic scatter and uncertainties in $M_{\rm BH}$, $R_{\rm b}$ and
$\sigma$. The best-fitting linear relations from the {\sc bces} and
\mbox{\sc linmix\_err} methods are consistent with each other within
their 1$\sigma$ uncertainties.  In this work we only focus on the {\sc
  bces} bisector $M_{\rm BH}-R_{\rm b}$ and $M_{\rm BH}-\sigma$
relations to allow a direct comparison with published works in
literature (e.g.,
\citealt{2013ApJ...764..151G,2019ApJ...887...10S}),Table~\ref{Table1}.
We also give in Table~\ref{Table1} the intrinsic scatter from the {\sc
  linmix\_err} regression fits, and the Spearman and Pearson
correlation coefficients\footnote{We note that the Spearman and
  Pearson correlation coefficients do not take into account the errors
  on the data points.} ($r_{\rm s}$ and $r_{\rm p}$, respectively)
together with the associated probabilities ($P$).

Directly  measured SMBH masses
are not available for the bulk of the galaxies (27/51) shown in
Fig.~\ref{Fig3}, hence  we opted  to fit the symmetric Ordinary
Least-Squares ({\sc ols}) bisector \citep{1992ApJ...397...55F} regression to
the (\mbox{$R_{\rm b}, M_{\rm BH}$}) data set, ignoring the
errors on $R_{\rm b}$ and $M_{\rm BH}$.

\subsection{$M_{\rm BH}-R_{\rm b}$ versus $M_{\rm BH}-\sigma$ relations
  for core-S\'ersic galaxies}\label{Sec3.2}

In Fig.~\ref{Fig1}, we show the $M_{\rm BH}-R_{\rm b}$ and
$M_{\rm BH}-\sigma$ diagrams for the 24 core-S\'ersic galaxies with
dynamically determined SMBH masses and with carefully measured
core-S\'ersic break radii. The masses of SMBHs correlate very strongly
with the core-S\'ersic break radii ($r_{\rm s} \sim 0.90$ and
$r_{\rm p} \sim 0.86$). The updated $M_{\rm BH}-R_{\rm b}$ relation
(Fig.~\ref{Fig1} and Table~\ref{Table1}) is in excellent agreement
with that from \citet[][his Table~5]{2019ApJ...886...80D} defined by
the subsample of 11 core-S\'ersic galaxies with measured $M_{\rm
  BH}$. Despite the discrepancy between the break radii obtained by us
and \citet{2013AJ....146..160R}, Section \ref{Sec_2_data}, our
\mbox{$M_{\rm BH}-R_{\rm b}$} relation is fully consistent with the
relations reported by \citet{2013AJ....146..160R} and \citet[their
Fig.~4]{2016Natur.532..340T}. However, the correlation for our
($M_{\rm BH},R_{\rm b}$) data set ($r_{\rm s} \sim 0.90$) is stronger
than that found by
\citet[][$r_{\rm s} \sim 0.77$]{2013AJ....146..160R} for their data
set.  Using the \lq\lq cusp radius'', i.e., the radius where the
negative logarithmic slope of the fitted Nuker model equals 1/2
($r_{{\gamma}^{'}=1/2}$) as a measure of the core size
\citep{1997ApJ...481..710C}, \citet[their eq.\
24]{2007ApJ...662..808L} reported $M_{\rm BH}-r_{{\gamma}^{'}}$
relation for their 11 core galaxies with directly measured SMBH
masses. Not only is the level of scatter in our
\mbox{$M_{\rm BH}-R_{\rm b}$} relation (Fig.~\ref{Fig1}) substantially
lower than that in the \mbox{$M_{\rm BH}-r_{{\gamma}^{'}}$}
relation\footnote{\citet{2007ApJ...662..808L} did not quote the
  scatter for their \mbox{$M_{\rm BH}-r_{{\gamma}^{'}}$} relation,
  thus our assessment is based on visual inspection of their Fig.~7.}
\citep[][their Fig.~7]{2007ApJ...662..808L}, but the intercepts from
the two relations also differ by $\sim 2\sigma$.

In Fig.~\ref{Fig1}, it is immediately apparent that the large-core
spheroids (filled purple boxes) {\it obey the log-linear
  $M_{\rm BH}-R_{\rm b}$ relation} established by the relatively less
massive, normal-core spheroids (filled red circles, blue disk symbol
and filled red stars), see Table~\ref{Table1}. The
$M_{\rm BH}-R_{\rm b}$ relation has an intrinsic scatter of
\mbox{$\epsilon \sim$0.33 dex} and an rms vertical scatter in the
\mbox{log $M_{\rm BH}$} direction of
\mbox{$\Delta_{\rm rms} \sim 0.29$ dex}. This can be compared to the
relatively weaker core-S\'ersic $M_{\rm BH}-\sigma$ relation with
$r_{\rm s} \sim 0.77$, $r_{\rm p} \sim 0.75$ and 62\% more scatter in
\mbox{log $M_{\rm BH}$} (\mbox{$\Delta_{\rm rms} \sim 0.47$ dex},
Table~\ref{Table1}), despite the two relations exhibiting a similar
level of intrinsic scatter ($\epsilon \sim 0.30 \pm
0.10$). \citet{2019ApJ...887..195M} measured a SMBH mass of
($4.0 \pm 0.80) \times 10^{10} M_{\sun}$ at the center of the BCG Holm
15A, the most massive dynamically determined black hole in the local
universe to date. Plotting this galaxy in Fig.~\ref{Fig1} using
data\footnote{We convert the circular break radii from
  \citet{2019ApJ...887..195M} into semi-major axis radii using the
  galaxy's ellipticity value at the break radius.} from
\citet[][$R_{\rm b } \sim 2.8 \pm 0.06 $ kpc,
$\sigma \sim 346 \pm 12.5$ km s$^{-1}$, see their
Fig.~11]{2019ApJ...887..195M}, extends the narrow
$M_{\rm BH}-R_{\rm b}$ sequence traced by other core-S\'ersic galaxies
to higher $M_{\rm BH}$ and $R_{\rm b}$, by a factor of $\sim2$ than
previously possible.  When we include Holm 15A and rerun the {\sc
  bces} bisector regression analysis on the 25 core-S\'ersic galaxies,
the resulting $M_{\rm BH}-R_{\rm b}$ relation (slope
$\sim 1.19 \pm 0.11$ and intercept at $R_{\rm b} = 250$ pc
$\sim 9.52 \pm 0.05$) is nearly identical to that for the 24
core-S\'ersic galaxies (see Table~\ref{Table1}), enforcing the
conclusion noted above.  However, the nature of the depleted core in
Holm 15A is controversial. \citet{2019ApJ...887..195M} fit the 2D
\mbox{core-S\'ersic}+S\'ersic+GaussianRing3D model to the Wendelstein
image of the BCG Holm 15A, finding a core size
\mbox{$R_{\rm b } \sim 2.8$ kpc}.  This structural analysis was
supplemented with their orbit analysis for the galaxy based on MUSE
spectroscopic data. However,
\citet{2015ApJ...807..136B,2016ApJ...819...50M} did not identify a
depleted core in their analyses of the galaxy's stellar light
distributions.

 \subsection{Steepening at the high-mass end of the $M_{\rm
     BH}-\sigma$ relation}\label{Sec3.3}

 In Fig.~\ref{Fig2}, we show our symmetric {\sc bces} bisector fit
 (dashed line) to the ($M_{\rm BH}, \sigma$) data for the composite
 sample of 141 galaxies---121 galaxies from
 \citet{2016ApJ...831..134V} and 20 normal-core galaxies
 (\citealt{2013AJ....146..160R}; \citealt[][and references
 therein]{2019ApJ...886...80D}), see Table~\ref{Table1}. The shaded
 region marks the 1$\sigma$ uncertainty for the fit. We find that
 S\'ersic and normal-core core-S\'ersic galaxies define a single
 log-linear $M_{\rm BH}-\sigma$ relation with a slope of
 $4.88 \pm 0.29$, \mbox{$\epsilon \sim $0.39 dex} and
 \mbox{$\Delta_{\rm rms} \sim 0.47$ dex} in the \mbox{log
   $M_{\rm BH}$}.  Excluding the 18 normal-core galaxies from our
 regression analysis, we find that the best-fitting {\sc bces}
 bisector $M_{\rm BH}-\sigma$ relation for the 121 galaxies
 \citep{2016ApJ...831..134V} remains roughly unchanged (slope
 $\sim 4.66 \pm 0.29$ and intercept $\sim 8.33 \pm 0.04$), albeit a
 slightly shallower slope. As noted above, a small fraction of the 121
 galaxies from \citet{2016ApJ...831..134V} with
 $M_{\rm BH }\ga 2 \times 10^{9} M_{\sun}$ and $\sigma \ga 280 $ km
 s$^{-1}$ are probably normal-core galaxies. If true, this further
 reinforces our conclusions.  Our $M_{\rm BH}-\sigma$ relation for
 S\'ersic and normal-core galaxies is in fair agreement with that from
 \citet{2016ApJ...831..134V} $M_{\rm BH}-\sigma$ with a slope of 5.35
 $\pm$ 0.23 reported for the sample of 230 galaxies used in their
 regression analysis (solid blue line).
 
 A trend emerges when we plot 13 large-core galaxies with $V$-band
 absolute magnitudes $M_{V} \la -23.50 \pm 0.10$ mag and stellar
 masses $M_{*} \ga 10^{12}M_{\sun}$: 3 with directly measured SMBH
 masses (purple boxes) and 10 with SMBH masses predicted using the
 tight \mbox{$M_{\rm BH}-R_{\rm b}$} relation (purple triangles),
 Fig.~\ref{Fig1} and Table~\ref{Table2}. Large-core galaxies are {\it
   offset upward} by $2.5-4\sigma_{\rm s}$ from the
 \mbox{$M_{\rm BH}-\sigma$} sequence traced by S\'ersic and
 normal-core galaxies (Fig.~\ref{Fig2}). This is indeed the case for
 Holm 15A (filled hexagon), which lies $\sim 1.1$ dex above the
 \mbox{non-(large-core)} \mbox{$M_{\rm BH}-\sigma$} relation if we
 consider it as a large-core galaxy candidate
 \citep{2019ApJ...887..195M}. We adopt that 1$\sigma_{\rm s}$ equals
 the intrinsic scatter $\epsilon= 0.39$ dex.  The $M_{\rm BH}-\sigma$
 relation defined by the S\'ersic and normal-core galaxies
 substantially underpredicts the SMBH masses for large-core
 galaxies. The inclusion of large-core galaxies in the regression
 analyses steepens the slopes of the $M_{\rm BH}-\sigma$ relations
 (Fig.~\ref{Fig2}, solid black line), regardless of the choice of
 linear regression method (see Table~\ref{Table1}).  Aside from the
 large-core galaxies, the only dramatic outlier in the
 $M_{\rm BH}-\sigma$ diagram is the flocculent spiral galaxy NGC~5055
 (Fig.~\ref{Fig2}). For reference, earlier studies have also reported
 brightest group and cluster galaxies would depart upward from the
 mean $M_{\rm BH}-\sigma$ relations (e.g.,
 \citealt{2006MNRAS.369.1081B,2007ApJ...662..808L,
2011Natur.480..215M,2012ApJ...756..179M,2012MNRAS.424..224H,2013ApJ...768...29V,2016Natur.532..340T,2018MNRAS.474.1342M}).
 This work (see also \citealt{2019ApJ...886...80D}) has provided
 detailed characterisation of the $M_{\rm BH}-\sigma$ relations and
 the brightest galaxies causing the offset at the high-mass end (e.g.,
 $R_{\rm b}$, $M_{V}$, $M_{\rm BH}$, $M_{\rm def}$ and $R_{\rm e}$ of
 large-core galaxies).

 Fitting the core-S\'ersic galaxies separately results in
 $M_{\rm BH}-\sigma$ relations with steeper slopes than those for the
 combined S\'ersic plus core-S\'ersic types (Table~\ref{Table1} and
 Figs.~\ref{Fig1} and \ref{Fig2}).  \citet{2013ApJ...764..184M}
 reported two different $M_{\rm BH}-\sigma$ relations for `core' and
 `power-law' galaxies have similar slopes but significantly different
 intercepts. Of their 28 core galaxies, three galaxies with low
 $\sigma$ values $\la 190$ km s$^{-1}$ (NGG~1374, NGG~4473 and
 NGG~5576) were reclassified as a S\'ersic type by
 \citet{2014MNRAS.444.2700D}. They also consider the potentially
 S\'ersic galaxy NGC~5128, with a \mbox{low $\sigma$} ($\sim 150$ km
 s$^{-1}$) and a strong nuclear dust, as core type. It appears that
 these four \mbox{low-$\sigma$} galaxies and the large-core galaxies
 in \citet{2013ApJ...764..184M} sample collectively lead to higher
 intercept for the core galaxies. Dividing their sample galaxies into
 S\'ersic and core-S\'ersic galaxies, \cite{2019ApJ...887...10S} also
 advocated two $M_{\rm BH}-\sigma$ relations with distinct slopes.  We
 find that the slopes for our core-S\'ersic $M_{\rm BH}-\sigma$
 relations are poorly constrained due to the fitted narrow baselines
 in $\sigma$. This has caused the slopes of the $M_{\rm BH}-\sigma$
 relations for normal-core galaxies and S\'ersic+normal-core galaxies
 to agree with $1\sigma$ overlapping error bars.  The marked division
 between S\'ersic and normal-core galaxies reported by
 \cite{2019ApJ...887...10S} is less evident in our work.  Importantly
 and as noted above, there is a full consistency between the
 $M_{\rm BH}-\sigma$ relation defined by S\'ersic galaxies and that
 established by the \mbox{S\'ersic plus normal-core} galaxy sample
 (e.g.,
 \citealt{2013ApJ...764..151G,2015MNRAS.446.2330S,2016ApJ...818...47S,2016ApJ...831..134V,2018MNRAS.473.5237K,2018MNRAS.477.3030K}).

\subsection{Offset in the $M_{\rm BH}-\sigma$  and $R_{\rm b}-M_{\rm BH}$ diagrams }\label{Sec3.4}

As noted in the introduction, the $M_{\rm BH}-\sigma$ relation
predicts that SMBH masses for the most massive galaxies (i.e.,
\mbox{$\sigma \sim 300 -390$ km s$^{-1}$},
\citealt{2007ApJ...662..808L,2007AJ....133.1741B}) cannot exceed
$M_{\rm BH} \sim 5 \times 10^{9} M_{\sun}$. However, Fig.~\ref{Fig2}
reveals four galaxies with $M_{\rm BH} \ga 10^{10} M_{\sun}$.  Also,
while the large break radii of all the three (four, if we include
\mbox{Holm 15A}) large-core galaxies with measured $M_{\rm BH}$
(Fig.~\ref{Fig1}), which constitute $12-17$\% of all the core-S\'ersic
galaxies with measured $M_{\rm BH}$ to date, are fully consistent with
the galaxies' large black hole masses, these are not predicted by the
$M_{\rm BH}-\sigma$ relation (Fig.~\ref{Fig2}).

Furthermore, the $R_{\rm b}-\sigma$ relation for normal-core galaxies
\citep[][their Fig.~5 and Table 3]{2014MNRAS.444.2700D} predicts that
break radii for the most massive galaxies are $R_{\rm b} \la 0.5$ kpc,
thus large-core galaxies ($R_{\rm b} > 0.5$ kpc) are not high-mass
extensions of the less massive normal-core galaxies.  In other words,
the $R_{\rm b}-\sigma$ relation for a sample of core-S\'ersic galaxies
containing large- and normal-core galaxies has a break occurring at
$R_{\rm b} \sim 0.5$ kpc.

Fig.~\ref{Fig3} shows the core-S\'ersic break radius versus SMBH mass
for our full sample of 51 core-S\'ersic galaxies; 24 have direct SMBH
masses (enclosed in boxes) and for the remaining 27 the SMBH masses
are based on $\sigma$ (Fig.~\ref{Fig3}a) and $L_{\rm sph}$
(Fig.~\ref{Fig3}b). The {\sc ols} bisector fit to the 41 core-S\'ersic
galaxies in Fig.~\ref{Fig3}(a)---i.e., excluding the 10
\mbox{large-core} galaxies with $\sigma$-based SMBH masses (see
Table~\ref{Table2})---yields \mbox{log ($R_{\rm b}/$pc) =
  (0.78$\pm$0.08) log ($M_{\rm BH}/2\times10^{9}$)} + (2.24
$\pm$0.04). This relation is entirely consistent with that shown in
Fig.~\ref{Fig1}(a) based on the galaxy sample with measured
$M_{\rm BH}$ (dotted lines), but significantly different from the
$\sigma$-based $R_{\rm b}-M_{\rm BH}$ relation for the full sample of
51 core-S\'ersic galaxies with a slope of $\sim 1.11\pm 0.11$ and an
intercept $\sim 2.54 \pm 0.10$. As noted in Section~\ref{Sec3.3}, the
discrepancy arises because the $M_{\rm BH}-\sigma$ relation for
S\'ersic and normal-core core-S\'ersic galaxies tends to underestimate
the true SMBH masses in extremely massive galaxies.

A similar tendency is exhibited in Fig.~\ref{Fig3}(b) for
$L_{\rm sph}$-based $M_{\rm BH}$; the
$R_{\rm b}-M_{{\rm BH}, L_{\rm sph}-{\rm based}}$ relation, which has
a slope of $\sim 0.86 \pm 0.07$ and an intercept $\sim 2.04 \pm 0.05$
for the 41 core-S\'ersic galaxies steepens for the full sample (slope
$\sim 1.03 \pm 0.08$ and intercept $\sim 2.12 \pm 0.09$). Nonetheless,
the $M_{\rm BH}-L$ relation is a better predictor of $M_{\rm BH}$ than
the $M_{\rm BH}-\sigma$ relation for large-core galaxies, confirming
past findings
(\citealt{2007AJ....133.1741B,2007ApJ...662..808L,2011Natur.480..215M,2012ApJ...756..179M,2013ApJ...768...29V,2018MNRAS.474.1342M}).

The offset nature of large-core galaxies in the
\mbox{$M_{\rm BH}-\sigma$} and \mbox{$R_{\rm b}-M_{\rm BH}$} diagrams
(Figs.~~\ref{Fig2}, \ref{Fig3}a and \ref{Fig3}b) is also revealed when
we compare the central stellar mass deficits ($M_{\rm def}$) of the
large-core galaxies with their \mbox{$\sigma$-, $L_{\rm sph}$- and
  $R_{\rm b}$-based} SMBH masses.  Table~\ref{Table2} lists the
(central stellar mass deficit)-to-(SMBH mass) ratios for our
large-core galaxies, where
$M_{\rm def} /M_{\rm BH,\sigma-{\rm based}} \sim$ 10 $-$ 160,
$M_{\rm def}/M_{{\rm BH}, L_{\rm sph}-{\rm based}} \sim$ 2 $-$ 70 and
$M_{\rm def}/M_{{\rm BH}, R_{\rm b}-{\rm based}} \sim$ 1 $-$ 5. The
prediction from high-accuracy \mbox{$N$-body simulations} of ``core
scouring'' by inspiraling binary SMBH (\citealt{2006ApJ...648..976M})
is that, after $\mathcal{N}$ successive dry major mergers, the
accumulated stellar mass deficit
$M_{\rm def} \approx 0.5 \mathcal{N}M_{\rm BH}$, where $M_{\rm BH}$ is
the final mass of the SMBH. Therefore, the $M_{\rm def}/M_{\rm BH}$
ratio is used as a proxy for the merger history of core-S\'ersic
galaxies. We computed $M_{\rm def}$ for the large-core galaxies by
measuring the central stellar luminosity deficit as the difference in
luminosity between inwardly-extrapolated outer S\'ersic profile of the
complete core-S\'ersic model and the core-S\'ersic model
\citep{2019ApJ...886...80D}. These luminosity deficits are then
converted into $M_{\rm def} $ using the galaxy stellar mass-to-light
ratios. The inferred merger rates from
$M_{\rm def}/M_{\rm BH,\sigma-{\rm based}}$ and
$M_{\rm def}/M_{{\rm BH}, L_{\rm sph}-{\rm based}}$ translate to
excessive number of major dry mergers ($\sim 5-320$) for $\sim$70\% of
the large-core spheroids.  Conversely,
$M_{\rm def}/M_{{\rm BH}, R_{\rm b}-{\rm based}}$ and
$M_{\rm def}/M_{{\rm BH,direct}}$ (Table~\ref{Table2}) correspond to
large-core galaxy formation via a reasonable number of (3$-$10) `dry'
major mergers, in agreement with observations measuring the close pair
fraction (e.g.,
\citealt{2004ApJ...608..752B,2006ApJ...640..241B,2012ApJ...744...85M,2013MNRAS.433..825L,2014MNRAS.445.1157C,2015MNRAS.449...49R})
and expectations from hierarchical structure formation models
(\citealt{2002MNRAS.336L..61H}, their Fig.~2;
\citealt{2009MNRAS.397..506K,2015MNRAS.449...49R}).

The excess merger rates derived above for large-core galaxies are too
large to be explained by additional core scouring mechanisms (e.g.,
repeated core-passage by a gravitational radiation kicked SMBH,
\citealt[][their Sections 5.3 and 5.4]{2014MNRAS.444.2700D}) which
have been suggested to generate large $M_{\rm def}/M_{\rm BH} \sim 5$
by amplifying a pre-existing depleted core carved out by a binary SMBH
(e.g.,
\citealt{1989ComAp..14..165R,2004ApJ...607L...9M,2012ApJ...744...74G}),
albeit see \citet{2020arXiv201104663N}.  Numerical simulations of a
Virgo-like galaxy cluster incorporating the effect of AGN feedback
\citep{2012MNRAS.420.2859M} produce a BCG with a core that is
extremely large in size ($\sim$10 kpc). While such extreme core
depletion by the AGN implies a marked reduction of the inferred merger
rate for the large-core galaxies, the simulated cores
\citep{2012MNRAS.420.2859M} are more than an order of magnitude larger
than the real cores observed in the Virgo cluster galaxies
\citep{2014MNRAS.444.2700D,2019ApJ...886...80D}: $R_{\rm b}\sim 0.11$
kpc for the BCG in the Virgo B subcluster (NGC 4472,
\citealt{2012ApJS..200....4F}) and $R_{\rm b} \sim 0.64$ kpc for the
Virgo A subcluster BCG (M87). For reference, the largest core size
measured in any real galaxy to date is $R_{\rm b} \sim 4.2$ kpc
\citep{2017MNRAS.471.2321D}. Furthermore, there are observed evidences
for core scouring by SMBH binaries.  The excess of tangential orbits
observed in the galaxy cores (e.g.,
\citealt{2003ApJ...583...92G,2014ApJ...782...39T,2016Natur.532..340T})
can naturally form when a SMBH binary decays orbitally, preferentially
ejecting stars that are on radial orbits
(\citealt{1997NewA....2..533Q, 2001ApJ...563...34M}).  Also,
\citet{2016Natur.532..340T,2018ApJ...864..113R,2019ApJ...887..195M}
revealed a strong correlation between the size of the depleted core in
a galaxy and the radius of the black hole'€™s sphere of
influence.

\subsection{Pathways for the growth of spheroids and black holes  in large-core,
 normal-core and  S\'ersic galaxies}\label{Bhgrowth}

As noted in the Introduction, depleted cores are thought to be
generated by inspiraling binary SMBHs created in major dry mergers
(\citealt{1980Natur.287..307B}). Supporting this scenario, we find
that the core size ($R_{\rm b}$) for large-core galaxies
($R_{\rm b} > 0.5$ kpc) and normal-core galaxies ($R_{\rm b} < 0.5$
kpc) is strongly correlated with the SMBH mass $M_{\rm BH}$
(\citealt{2007ApJ...662..808L,2012ApJ...755..163D,2013AJ....146..160R,2014MNRAS.444.2700D,2016Natur.532..340T,2019ApJ...887..195M,2019ApJ...886...80D}).

We emphasized above that S\'ersic and normal-core galaxies appear to
follow a single log-linear $M_{\rm BH}-\sigma$ relation, while
large-core galaxies, the bulk ($\sim$77\%) of which are BCGs, are
offset (to higher $M_{\rm BH}$), Table~\ref{Table1} and
Figs.~\ref{Fig1} and \ref{Fig2}. When core-S\'ersic galaxies are
fitted separately, the resulting core-S\'ersic $M_{\rm BH}-\sigma$
relation, with a steeper slope than that for the combined S\'ersic
plus core-S\'ersic types, is poorly
constrained. \citet{2019MNRAS.487.4827K} argue that present-day
galaxies, with overmassive black holes that offset upward from the
$M_{\rm BH}-\sigma$ relation, may be descendants of the compact blue
nuggets formed at $z \ga 6$, possibly connecting high-redshift blue
nuggets to large-core galaxies. Taken together our findings are
consistent with the two breaks recently detected in $\sigma-L_{V}$
relation occurring at \mbox{$M_{V} \sim -21.0$ mag} and
\mbox{$M_{V} \sim -23.5$ mag} \citep[][his
Fig.~7]{2019ApJ...886...80D}, which coincide with the S\'ersic versus
normal-core and normal-core versus large-core divides,
respectively. Akin ot this, we find that the $R_{\rm b}-\sigma$
relation \citep[][their Fig.~5 and Table 3]{2014MNRAS.444.2700D} has a
break occurring at \mbox{$R_{\rm b} \sim 0.5$ kpc}. Our findings carry
important implications for the formation origins of large- and
normal-core spheroids.

The offset tendency of large-core spheroids
(\mbox{$M_{V} \la -23.5 \pm 0.10$ mag}) in the $M_{\rm BH}-\sigma$ and
$R_{\rm b}-\sigma$ diagrams suggests that these galaxies are built via
major `dry' mergers, that create their large cores and grow their
black hole masses and spheroid stellar masses, while keeping their
velocity dispersion relatively unaffected (e.g.,
\citealt{2003MNRAS.342..501N,2007ApJ...658...65C,
  2012ApJ...744...63O,2013MNRAS.429.2924H}). In
\citet{2019ApJ...886...80D} we find that large-core spheroids are more
likely to experience a higher proportion of major mergers than
normal-core spheroids, consistent with the expectation that BCGs
experience a more intense merging and accretion events than galaxies
with relatively low luminosities (e.g.,
\citealt{,2006MNRAS.369.1081B,2007MNRAS.375....2D,2007AJ....133.1741B,2013MNRAS.435..901L}).
We \citep{2019ApJ...886...80D} revealed a flattening of the slope of
the $\sigma-L_{V}$ relation at the most luminous end, where the
velocity dispersion of the large-core spheroids appears to saturate,
contrary the velocity dispersion for normal-core galaxies, which
increases with galaxy luminosity as
$\sigma \propto L^{1/(3.50 \pm 0.61)}$. This observed trend of the
$\sigma-L_{V}$ relation for the most massive galaxies was recovered in
the theoretical models of \citet[][see their
Fig.~5]{2016MNRAS.459.4109T}.

The $\sigma-L_{V}$ relation for normal-core galaxies
($-21.0 \ga M_{V} \ga -23.5$ mag) may suggest that these galaxies
evolve through a few (1$-$8) successive gas-poor (but not purely
`dry') major\footnote{\citet{2013ApJ...769L...5K} argue that a few
  dissipationless mergers between two nearly equal-mass galaxies
  involving SMBHs are a viable formation mechanism for core-S\'ersic
  galaxies rather than a series of several minor mergers.}  mergers
since $z \sim 2$
\citep[e.g.,][]{2004ApJ...608..752B,2006ApJ...640..241B,2012ApJ...744...85M,2012ApJ...755..163D,2013ApJ...768...36D,2014MNRAS.444.2700D,2015ApJ...798...55D,2018MNRAS.475.4670D},
accompanied by low level star formation detected by {\it GALEX}
\citep[e.g.,][]{2007ApJS..173..185G,2018ApJS..234...18B}. That is,
normal-core galaxies are `red but not strictly dead', see \citet
{2011MNRAS.418L..74D,2019MNRAS.484.4413H,2019MNRAS.486.1404D}. We
cannot currently draw a firm conclusion, but this issue will be
further investigated in a forthcoming paper (Dullo 2020, in prep.). On
the other hand, S\'ersic galaxies ($M_{V} \ga -21.0$ mag) are said to
be products gas-rich major mergers and (major merger)-free processes
(e.g., secular processes and minor mergers),
\citep[e.g.,][]{1997AJ....114.1771F,2009ApJS..181..135H,2019ApJ...871....9D}. Gas-rich
processes increase a galaxy's velocity dispersion, black hole mass and
spheroid stellar mass, explaining the steepened $\sigma-L_{V}$
relation for S\'ersic galaxies (see
\citealt{2013ApJ...769L...5K,2018MNRAS.473.5237K,2018MNRAS.477.3030K,2019ApJ...887...10S}
; \citealt[][his Fig.~7]{2019ApJ...886...80D}).

\section{Conclusions}\label{Conc} 

Bright early-type (core-S\'ersic) galaxies have a partially depleted
core, a flattening in their inner stellar light distributions relative
to the inward extrapolation of the spheroid's outer S\'ersic profile,
thought to be created by coalescing binary SMBHs. The core-S\'ersic
model, which describes the light profiles of the spheroidal components
of core-S\'ersic galaxies, enables the flattened core sizes to be
measured by its break radius ($R_{\rm b}$).  We probed how well the
mass of SMBHs ($M_{\rm BH}$) correlates with the central velocity
dispersion ($\sigma$) and core size ($R_{\rm b}$) of their host
galaxies, placing a strong emphasis on the high-mass end. We do so
using a large sample of 144 galaxies with dynamically determined SMBH
masses, of which 24 are core-S\'ersic galaxies robustly identified
based on the analysis of the galaxy high-resolution {\it HST}
imaging. \citet{2013AJ....146..160R} have investigated the
$M_{\rm BH}- R_{\rm b}$ relation using their dataset for 23
core-S\'ersic galaxies, 21 of which are in common with our sample. We
acknowledge that there may be more core-S\'ersic galaxies with
measured SMBH mass in our sample, but even if this is true our
conclusions remain unchanged. Our principal conclusions are:\\

1) SMBH mass and core size correlate remarkably well.  This tight
$M_{\rm BH}- R_{\rm b}$ relation, which is stronger than the
$M_{\rm BH}- \sigma$ relation. In line with previous works (e.g.,
\citealt{2016Natur.532..340T}), we established the
$M_{\rm BH}- R_{\rm b}$ relation out to the high-mass end. The
relation \mbox{$M_{\rm BH} \propto R_{\rm b}^{1.20 \pm 0.14}$} is
defined by core-S\'ersic (i.e., combination of normal-core plus
large-core) galaxies with $M_{\rm V} \la -21$ mag and it has
$r \sim 0.90$, a vertical rms scatter in the \mbox{log $M_{\rm BH}$}
of $\sim 0.29$ dex and an intrinsic scatter of $0.33 \pm 0.07$ dex. We
also checked and found that the BCG Holm 15A, which hosts the most
massive dynamically measured the black hole in the local universe to
date ($4.0 \pm 0.80 \times 10^{10} M_{\sun}$,
\citealt{2019ApJ...887..195M}) extends the tight
$M_{\rm BH}- R_{\rm b}$ sequence traced by other core-S\'ersic
galaxies to high $M_{\rm BH}$ and $R_{\rm b}$, by a factor of $\sim 2$
than previously possible. However, this is if the galaxy has a
depleted core (\mbox{$R_{\rm b } \sim 2.8$ kpc},
\citealt{2019ApJ...887..195M}), although
\citet{2015ApJ...807..136B,2016ApJ...819...50M} did not identify one.

2) Separating our sample into S\'ersic, normal-core and large-core
galaxies, we find that S\'ersic and normal-core galaxies unite to
define a single log-linear $M_{\rm BH}-\sigma$ relation with a slope
of $4.88 \pm 0.29$, \mbox{$\epsilon \sim $0.39 dex} and
\mbox{$\Delta_{\rm rms} \sim 0.47$ dex} in the \mbox{log
  $M_{\rm BH}$}.  A key result is that large-core galaxies (four of
which, including Holm 15A, have measured $M_{\rm BH}$, while
$M_{\rm BH}$ for the remaining 10 is predicted using the
$M_{\rm BH}-R_{\rm b}$ relation) are offset upward systematically by
$(2.5-4)\times\sigma_{\rm s}$ from the \mbox{$M_{\rm BH}-\sigma$}
sequence traced by S\'ersic and normal-core galaxies. While
core-S\'ersic galaxies alone seem to follow steeper
$M_{\rm BH}-\sigma$ relations than the combined S\'ersic plus
core-S\'ersic types, the former are poorly constrained due to the
narrow baseline in $\sigma$ probed by the core-S\'ersic galaxies.
Previous studies also argued that the $M_{\rm BH}-\sigma$ relation
gets steeper for the brightest group and cluster galaxies (e.g.,
\citealt{2006MNRAS.369.1081B,2007ApJ...662..808L,2011Natur.480..215M,2013ApJ...768...29V}).
Using four measured and 10 predicted $M_{\rm BH}$ for large-core
galaxies, we (see also \citealt{2019ApJ...886...80D}) have offered
insights into the possible nature of the galaxies deriving the offset
at the high-mass end (e.g., $R_{\rm b}$, $M_{V}$, $M_{\rm BH}$,
$M_{\rm def}$ and $R_{\rm e}$ of large-core galaxies). When S\'ersic,
normal-core and large-core galaxies are fitted together, the resulting
$M_{\rm BH}-\sigma$ relation steepens to give a slope of
$5.76 \pm 0.37$. We also find that the $R_{\rm b}-\sigma$ relation
\citep[][their Fig.~5 and Table 3]{2014MNRAS.444.2700D} has a break
occurring at \mbox{$R_{\rm b} \sim 0.5$ kpc}, which is internally
consistent with the normal-core versus large-core divide.

3) For core-S\'ersic galaxies, the $M_{\rm BH}- \sigma$ relation
exhibits $\sim$ 62\% more scatter in \mbox{log $M_{\rm BH}$} than the
$M_{\rm BH}- R_{\rm b}$ relation, favouring $R_{\rm b}$ to estimate
the ultramassive black hole masses ($M_{\rm BH} \ga 10^{10}M_{\sun}$)
in the most massive (large-core) spheroids with
$M_{V} \la -23.50 \pm 0.10$ mag and $M_{*} \ga 10^{12}M_{\sun}$,
instead of e.g., $\sigma$ (Figs.~\ref{Fig1}, \ref{Fig2} and
\ref{Fig3}).

4) Taken together our findings reveal that the $M_{\rm BH}-\sigma$
relation for the S\'ersic and normal-core galaxies substantially
underpredicts the SMBH masses of large-core galaxies, i.e., the
measured ones and those estimated by the $M_{\rm BH}-R_{\rm b}$
relation. The assumption in galaxy formation models involving SMBHs
that all classical bulges and elliptical galaxies obey a single
$M_{\rm BH}-\sigma$ relation may be invalid. Extremely massive
galaxies need to be treated separately in such models.

5) We argue that large-core spheroids are consequences of multiple
major `dry' merging events involving super/ultramassive BHs, that
create their large cores and simply add the black hole masses, stellar
masses and luminosities of their progenitors, while keeping the
velocity dispersion relatively unaffected. This conclusion is
consistent with the offset tendency of large-core galaxies in the
$M_{\rm BH}-\sigma$ diagrams and the flattening in the $\sigma-L_{V}$
relation observed at $M_{V} \la -23.5$ mag
\citep{2019ApJ...886...80D}.

We highlight that the high SMBH masses of large-core galaxies means
they are targets of great interest for the detection and study of
gravitational waves (GWs) using pulsar timing array (PTA) and {\it
  Laser Interferometer Space Antenna (LISA)}. Current PTAs are
sensitive to gravitational waves emitted by merging binary SMBHs with
total mass $\ga 2 \times 10^{9}M_{\sun}$ at a distance $D\la$ 200 Mpc
(see \citealt{2019A&ARv..27....5B}), while {\it LISA} is sensitive to
binary SMBHs with mass $ 10^{5} - 10^{10}M_{\sun}$
\citep{2017arXiv170200786A,2020MNRAS.491.2301K}.

\section{ACKNOWLEDGMENTS}

We thank the referee for their suggestions that improved the
paper. B.T.D acknowledges supports from a Spanish postdoctoral
fellowship `Ayudas 1265 para la atracci\'on del talento
investigador. Modalidad 2: j\'ovenes investigadores.' funded by
Comunidad de Madrid under grant number 2016-T2/TIC-2039 and from the
award of `Estancias movilidad en el extranjero 'Jos\'e Castillejo'
para j\'ovenes doctores 2019', grant number CAS19/00344, offered by
the Spanish ministry of science and innovation for a research stay at
Swinburne University of Technology (01/01/2020-31/03/2020).  B.T.D
acknowledges financial support from grant `Ayudas para la
realizaci\'on de proyectos de I+D para j\'ovenes doctores 2019.'
funded by Comunidad de Madrid and Universidad Complutense de Madrid
under grant number PR65/19-22417.  B.T.D and A.G.d.P acknowledge
financial support from the Spanish Ministry of Science, Innovation and
Universities (MCIUN) under grant numbers AYA2016-75808-R and
RTI2018-096188-B-I00.  J.H.K. acknowledges financial support from the
European Union's Horizon 2020 research and innovation programme under
Marie Sk\l odowska-Curie grant agreement No 721463 to the SUNDIAL ITN
network, from the State Research Agency (AEI-MCINN) of the Spanish
Ministry of Science and Innovation under the grant ``The structure and
evolution of galaxies and their central regions" with reference
PID2019-105602GB-I00/10.13039/501100011033, and from IAC project
P/300724, financed by the Ministry of Science and Innovation, through
the State Budget and by the Canary Islands Department of Economy,
Knowledge and Employment, through the Regional Budget of the
Autonomous Community.

\bibliographystyle{apj}
\bibliography{Bil_Paps_biblo.bib}

\label{lastpage}
\end{document}